\documentclass[conference]{IEEEtran}

\usepackage{amsmath}
\usepackage{amsfonts}
\usepackage{amssymb}
\usepackage{graphicx}

\title{Progressive Barrier Lyapunov Functions for Nonlinear Constrained Control Systems}
\author{\IEEEauthorblockN{Hamed Rahimi Nohooji}
\IEEEauthorblockA{\textit{Int. Centre for Security, Reliability and Trust (SnT)} \\
\textit{Automation and Robotics Research Group (ARG)}\\
\textit{University of Luxembourg}\\
Luxembourg \\
ORCID:0000-0001-9429-7164}
\and
\IEEEauthorblockN{Holger Voos}
\IEEEauthorblockA{\textit{Faculty of Science, Technology and Medicine (FSTM)} \\
\textit{Dept. of Engineering, University of Luxembourg}\\
\IEEEauthorblockA{\textit{Int. Centre for Security, Reliability and Trust (SnT)} \\
\textit{Automation Robotics Research Group, University of Luxembourg}\\
ORCID:0000-0002-9600-8386}}
}
\date{}

\begin{document}

\maketitle

\begin{abstract}
This paper introduces the Progressive Barrier Lyapunov Function (p-BLF) for output- and full-state-constrained nonlinear control systems. Unlike traditional BLF methods, where control effort continuously increases as the state approaches the constraint boundaries, the p-BLF maintains minimal control effort in unconstrained regions and increases it progressively toward the boundaries. In contrast to previous methods with predefined safe zones and abrupt control activation, the p-BLF provides a smooth transition, improving continuity in the control response and enhancing stability by reducing chattering. Two forms of the p-BLF, logarithmic-based and rational-based, are developed to handle systems with either output constraints or full-state constraints. Theoretical analysis guarantees that all system states remain within the defined constraints, ensuring boundedness and stability of the closed-loop system. Simulation results validate the effectiveness of the proposed method in constrained nonlinear control.

\end{abstract}


\section{Introduction}

In modern control systems, ensuring that state variables remain within predefined bounds is important for maintaining system stability and safety, particularly in safety-critical applications such as autonomous vehicles, robotics, and aerospace systems. These constraints are often required to avoid system failure, maintain safe operation, or ensure compliance with physical or operational limits. To address these constraints, Barrier Lyapunov Functions (BLFs) have emerged as a powerful tool, providing a systematic approach to enforce hard constraints on the system states \cite{tee2009control, tee2011control, nohooji2018neural}.

Traditional BLF-based control strategies have been extensively studied for enforcing state constraints in nonlinear systems. Various forms of BLFs have been proposed, including tangent-type \cite{rahimi2018neural2}, integral-type \cite{tang2016adaptive}, and logarithmic-type functions \cite{tee2009barrier, nohooji2020constrained}, all designed to increase rapidly as the state approaches the constraint boundary. While this property ensures strong constraint enforcement, it also leads to high control effort even when the system states remain far from the boundaries. Such unnecessary control activation can increase energy consumption and accelerate actuator wear, particularly in applications where the constraints are rarely active.

To overcome the high control effort associated with traditional BLFs, the Zone Barrier Lyapunov Function (zBLF) was introduced as an alternative that defines a predefined safe region around the origin \cite{liang2023zone, liang2023nonlinear}. Within this zone, system states are allowed to evolve freely without control activation, and corrective effort is applied only as the state approaches the zone boundary. This structure helps reduce unnecessary control actions, improve energy efficiency, and limit actuator usage. However, zBLF presents limitations related to its discontinuous control behavior at the zone boundary. The abrupt activation of control introduces nonsmoothness in the system's response, which can lead to discontinuities in the velocity and sharp variations in acceleration. These effects may cause chattering, degraded tracking performance, and increased mechanical stress in applications that require smooth and continuous dynamics. Moreover, the fixed zone definition complicates the design of controllers that aim to provide a gradual and continuous increase in effort as the constraint boundary is approached.

To address these challenges, we propose a new class of BLFs, referred to as the Progressive Barrier Lyapunov Function (p-BLF). The p-BLF method retains the low-effort behavior of zone-based approaches but removes the need for explicit zone definitions. It provides a gradual increase in control effort as the system state approaches the constraint boundaries, avoiding nonsmooth control activation. This progression ensures continuity in the Lyapunov derivative and improves stability by reducing chattering and discontinuities in the system response. The control effort remains minimal in unconstrained regions and increases only as needed to enforce the constraints.

The key contributions of this paper are as follows:
\begin{itemize}
    \item A new class of BLFs, called p-BLF, is introduced to enable gradual and continuous control activation near constraint boundaries, addressing nonsmoothness and discontinuities in existing methods.
    \item The p-BLF framework maintains minimal control effort in unconstrained regions and increases it progressively as the state approaches the constraints, improving control efficiency and smoothness.
    \item Theoretical analysis and simulations demonstrate that the proposed approach ensures constraint satisfaction, closed-loop stability, and continuous control performance.
\end{itemize}

The rest of the paper is structured as follows. Section II presents the problem formulation and relevant preliminaries. Section III introduces the proposed p-BLF, detailing its logarithmic-based and rational-based forms, and presents a control design example for a second-order SISO system. The section further extends the control design to higher-order systems with output and full-state constraints. Section IV presents simulation results, and Section V concludes the paper.

\section{Problem Formulation and Preliminaries}

\textbf{System Description:}  
We consider a class of nonlinear systems in strict-feedback form:

\begin{align} \label{eq:system}
\dot{x}_i &= f_i(x_1, \dots, x_i) + g_i(x_1, \dots, x_i) x_{i+1}, && i = 1, \dots, n-1, \nonumber \\
\dot{x}_n &= f_n(x_1, \dots, x_n) + g_n(x_1, \dots, x_n) u, \nonumber \\
y &= x_1.
\end{align}
where \( x_i \in \mathbb{R} \) represents the system states, \( u \in \mathbb{R} \) is the control input, and \( y = x_1 \) is the system output. The functions \( f_i(\cdot) \) and \( g_i(\cdot) \) are assumed to be known and smooth.

\textbf{Assumptions:}  
1. The desired trajectory \( x_d(t) \) and its derivatives \( x_{d}^{(i)}(t) \), \( i = 1, \dots, n \), are bounded by:
\begin{equation}
|x_d(t)| < k_{xd}, \quad |x_{d}^{(i)}(t)| < k_{xdi}, \quad i = 1, \dots, n.
\end{equation}
These bounds ensure that the reference signal remains within the specified constraints for all time.

\textbf{Lemma 1 (Barbalat’s Lemma):} 
Consider a differentiable function \( f(t) \). If the following conditions hold:
\begin{enumerate}
    \item The time derivative \( \dot{f}(t) \) is uniformly continuous.
    \item \( \lim_{t \to \infty} f(t) \) is finite.
\end{enumerate}
Then, \( \lim_{t \to \infty} \dot{f}(t) = 0 \).

\textbf{Control Objective:}  
Given the nonlinear system described by \eqref{eq:system}, the objective is to design a control law \( u(t) \) such that the output \( y(t) = x_1(t) \) tracks a bounded desired trajectory \( x_d(t) \), while satisfying the state constraints. Two scenarios are considered: (i) output constraint \( |x_1(t)| < k_{x1} \), and (ii) full-state constraints \( |x_i(t)| < k_{xi} \), \( i = 1, \dots, n \). To achieve this, a p-BLF is developed and integrated into the control design to ensure constraint satisfaction, closed-loop stability, and asymptotic tracking, while maintaining boundedness of all signals for \( t \geq 0 \).

\textbf{Barrier Lyapunov Functions} \cite{tee2009barrier} is a scalar function \( V(x) \), defined with respect to the system \( \dot{x} = f(x) \) on an open region \( D \) containing the origin. The function is continuous, positive definite, and has continuous first-order partial derivatives at every point of \( D \). Additionally, it holds that \( V(x) \to \infty \) as \( x \) approaches the boundary of \( D \). Along the solution of \( \dot{x} = f(x) \), for \( x(0) \in D \), it satisfies \( V(x(t)) \leq b \) for all \( t \geq 0 \), where \( b \) is some positive constant.

One example of a standard BLF is given by the logarithmic form \( V(x) = \frac{1}{2} \ln \left( \frac{k^2}{k^2 - x^2} \right) \), where \( k \) represents the constraint boundary.

\textbf{Lemma 2} \cite{tee2009control}\textbf{:} For any positive constants \( k_{i} \), \( i = 1, 2, \dots, n \), let \( Z := \{ z \in \mathbb{R}^n : |z_i| < k_{i}, i = 1, 2, \dots, n \} \subset \mathbb{R}^n \) be an open set. Consider the system:
\[
\dot{z} = h(t, z)
\]
where \( h : \mathbb{R}^+ \times \mathbb{N} \to \mathbb{R}^n \) is piecewise continuous in \( t \) and locally Lipschitz in \( z \), uniformly in \( t \), on \( \mathbb{R}^+ \times Z \). Define \( Z_i := \{ z_i \in \mathbb{R} : |z_i| < k_{i} \} \subset \mathbb{R} \). Suppose there exist positive definite functions \( V_i : Z_i \to \mathbb{R}^+ \) (for \( i = 1, 2, \dots, n \)), which are continuously differentiable on \( Z_i \) and satisfy
\[
V_i(z_i) \to \infty \quad \text{as} \quad z_i \to \pm k_{i}.
\]
Let \( V(z) := \sum_{i=1}^{n} V_i(z_i) \), and suppose \( z_i(0) \in Z_i \) for all \( i = 1, 2, \dots, n \). If the inequality:
\[
\dot{V} = \frac{\partial V}{\partial z} h \leq 0
\]
holds, then \( z_i(t) \in Z_i \) for all \( t \in [0, \infty) \).

\textbf{Zone Barrier Lyapunov Functions} \cite{liang2023zone} is a scalar function \( V(x) \), defined with respect to the system \( \dot{x} = f(x) \), where the state space is divided into zones. These zones consist of a free-motion zone, where the control effort is zero (\( u = 0 \)), and a boundary zone, where the control action increases as the system state approaches the constraint. The ZBLF is continuous, positive definite, and approaches infinity as the system state \( x \) reaches the boundary \( k \). The function enforces constraints by ensuring that control action is only applied near the boundary, allowing for zero control interventions in the interior zone, leading to reduced energy consumption and actuator wear when the system operates within the free-motion zone.

To the best of the authors' knowledge, the only ZBLF developed so far is the logarithmic form:
 \(
V(x) = \frac{1}{2} \ln \left( \frac{k^2 e^{-2b}}{k^2 - x^2} \right),
\)
where \( k \) defines the constraint boundary and \( b \) is a design parameter controlling the size of the free-motion zone \cite{liang2023zone, liang2023nonlinear}.

In this paper, we modify the ZBLF by introducing a smooth transition in the control effort, where the control remains near zero when the system state is far from the constraints and gradually increases as the state approaches the boundaries. This modification eliminates abrupt changes in control activation, providing smoother control transitions and enhancing system stability and performance.

\section{Control Design with Progressive Barrier Lyapunov Function}

\subsection{Progressive Barrier Lyapunov Function}

The \textbf{Progressive Barrier Lyapunov Function (p-BLF)} is introduced to enable smooth and continuous control effort modulation. It ensures minimal control activation near the origin and progressively increases the effort as the system state approaches the constraint boundary. In contrast to zone-based BLFs, which may cause abrupt control switching, the p-BLF provides a gradual transition, improving continuity and closed-loop stability.

For a system state \( x \) subject to the constraint \( |x| < k \), two example forms of the p-BLF are considered:

1. $ V(x) = \frac{1}{2\beta} \ln \left( \frac{k^2}{k^2 - x^2} \right) $

2. $ V(x) = \frac{x^2}{2(k^2 - x^2)(1 + \beta x^2)} $.

\noindent
In both examples, \( \beta > 1 \) is selected to ensure that the Lyapunov function \( V(x) \) exhibits a slower growth rate near the origin, allowing for minimal control effort when the system state is far from the constraint boundaries.

To simplify the notation, the dependence on state and time variables will be omitted from this point forward, whenever it can be done without causing confusion.

\subsection{Motivation Example: Control Design for a Second Order SISO System}

We consider a second-order nonlinear SISO system in the form:
\begin{equation}
\dot{x}_1 = f_1(x_1) + g_1(x_1) x_2
\label{eq:1}
\end{equation}
\begin{equation}
\dot{x}_2 = f_2(x_1, x_2) + g_2(x_1) u
\label{eq:2}
\end{equation}

\noindent
where \( u \) is the control input. Let \( z_1 = x_1 - x_d \) and \( z_2 = x_2 - \alpha_1 \), where \( \alpha_1 \) is a stabilizing function. The objective is to design an output constrained control law \( u \) that respects the state constraint \( |x_1(t)| < k_{x1} \) while maintaining system stability.
We illustrate both forms of the p-BLF presented in the previous section with examples.

\subsubsection{Logarithmic-Based p-BLF}

The logarithmic Lyapunov function is given by:
\begin{equation}
V_1 = \frac{1}{2\beta} \ln \left( \frac{k^2}{k^2 - z_1^2} \right)
\label{eq:3}
\end{equation}

\noindent
where $k=k_{x1}-k_{xd}$. The time derivative of \( V_1 \) is:
\begin{equation}
\dot{V}_1 = \frac{z_1 \dot{z}_1}{\beta (k^2 - z_1^2)}
\label{eq:4}
\end{equation}

\noindent
Substitute \( \dot{z}_1 \) using the system dynamics \eqref{eq:1}, \( \dot{V}_1 \) becomes

\begin{equation}
\dot{V}_1 = \frac{z_1 \left( f_1(x_1) + g_1(x_1) (z_2 + \alpha_1) - \dot{y}_d \right)}{\beta (k^2 - z_1^2)}.
\label{eq:6}
\end{equation}

\noindent
The stabilizing function \( \alpha_1 \) is designed as:

\begin{equation}
\alpha_1 = \frac{1}{g_1} \left( -f_1 - \beta \kappa_1 z_1 \left( k^2 - z_1^2 \right) + \dot{y_d} \right).
\label{eq:7}
\end{equation}

\noindent
Substituting \eqref{eq:7} into \( \dot{V}_1 \) gives

\begin{equation}
\dot{V}_1 = -\kappa_1 z_1^2 + \frac{g_1 z_1 z_2}{\beta (k^2 - z_1^2)}
\label{eq:8}
\end{equation}

We define the augmented Lyapunov function \( V_2 \) as:

\begin{equation}
V_2 = V_1 + \frac{1}{2} z_2^2.
\label{eq:9}
\end{equation}

\noindent
The time derivative of \( V_2 \) is given by
\begin{equation}
\dot{V}_2 = -\kappa_1 z_1^2 + \frac{g_1 z_1 z_2}{\beta (k^2 - z_1^2)} + z_2 \dot{z}_2
\label{eq:10}
\end{equation}

\noindent
Substitute \( \dot{z}_2 \) with the system dynamics \eqref{eq:2}, \( \dot{V}_2 \) becomes:
\begin{equation}
\dot{V}_2 = -\kappa_1 z_1^2 + \frac{g_1 z_1 z_2}{\beta (k^2 - z_1^2)} + z_2 \left( f_2 + g_2 u - \dot{\alpha_1} \right).
\label{eq:12}
\end{equation}

The control law \( u \) is designed as:

\begin{equation}
u = \frac{1}{g_2} \left( -f_2 + \dot{\alpha_1} - \kappa_2 z_2 - \frac{g_1 z_1}{\beta (k^2 - z_1^2)} \right),
\label{eq:13}
\end{equation}

\noindent
then, substituting this into \eqref{eq:12} gives
\begin{equation}
\dot{V}_2 = - \kappa_1 z_1^2 - \kappa_2 z_2^2.
\label{eq:14}
\end{equation}

This ensures \( \dot{V}_2 \leq 0 \), establishing the stability of the closed-loop system, and the boundedness of $x_1$ considering Assumption 1, and Lemma 2.

\subsubsection{Rational-Based p-BLF}

For the rational-based p-BLF, the Lyapunov function is defined as:

\begin{equation}
V_1 = \frac{z_1^2}{2(k^2 - z_1^2)(1 + \beta z_1^2)}
\label{eq:15}
\end{equation}

\noindent
The time derivative of \( V_1 \) is given by

\begin{equation}
\dot{V}_1 = \frac{(\beta z_1^4 + k^2)}{(k^2 - z_1^2)^2 (1 + \beta z_1^2)^2} z_1 \dot{z}_1.
\label{eq:16}
\end{equation}

\noindent
Define \( N = (\beta z_1^4 + k^2) \) and \( D = (k^2 - z_1^2)^2 (1 + \beta z_1^2)^2 \). The stabilizing function \( \alpha_1 \) is:

\begin{equation}
\alpha_1 = \frac{1}{g_1} \left( -f_1 - \kappa_1 \frac{N}{D} z_1 + \dot{y_d} \right).
\label{eq:17}
\end{equation}

\noindent
Substituting \( \alpha_1 \) and \( \dot{z}_1 \), \( \dot{V}_1 \) becomes:
\begin{equation}
\dot{V}_1 = - \kappa_1 z_1^2 + g_1 \frac{N}{D} z_1 z_2.
\label{eq:18}
\end{equation}

Define the augmented Lyapunov function \( V_2 \) as
\begin{equation}
V_2 = V_1 + \frac{1}{2} z_2^2.
\label{eq:19}
\end{equation}

The time derivative of \( V_2 \) is given by
\begin{equation}
\dot{V}_2 = - \kappa_1 z_1^2 + g_1 \frac{N}{D} z_1 z_2 + z_2 \left( f_2 + g_2 u - \dot{\alpha_1} \right).
\label{eq:20}
\end{equation}

The control law \( u \) is designed as:

\begin{equation}
u = \frac{1}{g_2} \left( -f_2 + \dot{\alpha_1} - \kappa_2 z_2 - g_1 \frac{N}{D} z_1 \right).
\label{eq:21}
\end{equation}

\noindent
Substituting \eqref{eq:21} into \eqref{eq:20} gives
\begin{equation}
\dot{V}_2 = - \kappa_1 z_1^2 - \kappa_2 z_2^2.
\label{eq:22}
\end{equation}

This ensures \( \dot{V}_2 \leq 0 \), establishing the stability of the closed-loop system, and the boundedness of $x_1$ considering Assumption 1, and Lemma 2 for the rational-based p-BLF.

\subsection{p-BLF for Higher-Order Nonlinear Systems with Output Constraints}

Backstepping is used in the control design in this section. Let $z_1 = x_1 - x_d$ and $z_i = x_i - \alpha_{i-1}$, for $i = 2, \dots, n$. Define the state constraint as \( |x_1(t)| < k_{x1} \), and let $k_1=k_{x1}-k_{xd}$ is the error constraint.  The Lyapunov function candidates are defined using the logarithmic-based p-BLF as follows:

\begin{equation}
\label{Lyp1}
    V_1 = \frac{1}{2 \beta} \ln \left( \frac{k_{1}^2}{k_{1}^2 - z_1^2} \right)
\end{equation}
\begin{equation}
\label{Lyp1i}
    V_i = V_{i-1} + \frac{1}{2} z_i^2, \quad i = 2, \dots, n
\end{equation}

\noindent
where $\beta > 0$ is a design parameter. The stabilizing functions and control law are designed as:

\begin{equation} \label{alpha1}
    \alpha_1 = \frac{1}{g_1} \left( -f_1 - \beta \kappa_1 z_1 \left( k_1^2 - z_1^2 \right) + \dot{y}_d \right)
\end{equation}
\begin{equation}  \label{alpha2}
    \alpha_2 = \frac{1}{g_2} \left( -f_2 + \dot{\alpha}_1 - \kappa_2 z_2 - \frac{g_1 z_1}{\beta \left( k_1^2 - z_1^2 \right)} \right)
\end{equation}
\begin{equation}  \label{alphai}
    \alpha_i = \frac{1}{g_i} \left( -f_i + \dot{\alpha}_{i-1} - \kappa_i z_i - g_{i-1} z_{i-1} \right), \quad i = 3, \dots, n
\end{equation}
\begin{equation}
\label{uout}
    u = \alpha_n
\end{equation}

\noindent
where $\kappa_i > 0$ are constants, and $\dot{\alpha}_{i-1}$ is given by:

\begin{equation}
    \dot{\alpha}_{i-1} = \sum_{j=1}^{i-1} \frac{\partial \alpha_{i-1}}{\partial x_j} \left( f_j + g_j x_{j+1} \right) + \sum_{j=0}^{i-1} \frac{\partial \alpha_{i-1}}{\partial y^{(j)}_d} y^{(j+1)}_d
\end{equation}

\noindent
Then, the closed-loop system dynamics are:

\begin{equation}
\label{zd1}
    \dot{z}_1 = -\beta \kappa_1 z_1 \left( k_1^2 - z_1^2 \right) + g_1 z_2
\end{equation}
\begin{equation}
    \dot{z}_2 = -\kappa_2 z_2 - \frac{g_1 z_1}{\beta \left( k_1^2 - z_1^2 \right)} + g_2 z_3
\end{equation}
\begin{equation}
    \dot{z}_i = -\kappa_i z_i - g_{i-1} z_{i-1} + g_i z_{i+1}, \quad i = 3, \dots, n-1
\end{equation}
\begin{equation}
\label{zdn}
    \dot{z}_n = -\kappa_n z_n - g_{n-1} z_{n-1}
\end{equation}

\textbf{Theorem 1:}  
Consider the nonlinear system described by \eqref{eq:1}, \eqref{eq:2} with stabilizing functions \eqref{alpha1}-\eqref{alphai}, and the control law \eqref{uout} under the Assumption 1, and assume the initial conditions satisfy \( |z_1(0)| < k_1 \). Let the design parameters \( \beta > 0 \) and \( \kappa_i > 0 \) be chosen appropriately. Then, the following properties hold:

(i) The signals \( z_i(t) \), \( i = 1, \dots, n \), remain in the compact set defined by:
   \[
   \mathcal{Z} = \left\{ z \in \mathbb{R}^n : |z_1| \leq D_{z_1}, \sum_{i=2}^{n} z_i^2 \leq 2V_n(0) \right\}
   \]
   where:
   \[
   D_{z_1} = k_1 \sqrt{1 - e^{-2\beta V_n(0)}}
   \]
   and \( V_n(0) \) is the initial value of the Lyapunov function.

(ii) The output \( x_1(t) \) remains within the constraint set \( |x_1(t)| < k_{x1} \) for all \( t \geq 0 \).

(iii) All closed-loop signals are bounded.

(iv) The tracking error \( z_1(t) = x_1(t) - x_d(t) \) converges to zero asymptotically as \( t \to \infty \), i.e., \( \lim_{t \to \infty} z_1(t) = 0 \).

\textbf{Proof:}

(i) From the closed-loop system dynamics \eqref{zd1}-\eqref{zdn} and the Lyapunov function defined by \eqref{Lyp1}-\eqref{Lyp1i}, the time derivative of \( V_n \) can be obtained \( \dot{V}_n = - \sum_{i=1}^{n} \kappa_i z_i^2 \), ensuring that \( \dot{V}_n \leq 0 \), which implies \[ V_n(t) \leq V_n(0) \] for all \( t \geq 0 \). Consequently, the signals \( z_i(t) \), \( i = 1, \dots, n \), remain bounded. Specifically, from \( \frac{1}{2\beta} \ln \left( \frac{k_1^2}{k_1^2 - z_1^2} \right) \leq V_n(0) \), it follows that \( |z_1(t)| \leq k_1 \sqrt{1 - e^{-2\beta V_n(0)}} \), and the remaining states satisfy \( \sum_{i=2}^{n} z_i^2 \leq 2 V_n(0) \). Therefore, the signals \( z_i(t) \), \( i = 1, \dots, n \), remain in the compact set \( \mathcal{Z} \).

(ii) From the fact that \( z_1(t) \leq D_{z_1} \), we have \( |x_1(t)| = |z_1(t) + x_d(t)| \leq D_{z_1} + |x_d(t)| \). Since \( |x_d(t)| \leq k_{xd} \), we conclude that:
\[
|x_1(t)| \leq k_1 + k_{xd} = k_{x1}
\]
Hence, the output constraint \( |x_1(t)| < k_{x1} \) is never violated.

(iii)
From (i), we know that the error signals \( z_1(t), \dots, z_n(t) \) are bounded. The boundedness of \( z_1(t) \) and \( x_d(t) \) implies that the state \( x_1(t) \) is bounded. Given that \( \dot{x}_d(t) \) is bounded by Assumption 1, it follows from \eqref{alpha1} that the stabilizing function \( \alpha_1(t) \) is also bounded. Consequently, since \( x_2 = z_2 + \alpha_1 \), the state \( x_2(t) \) is bounded. From  \eqref{alpha2}, \( \alpha_2(t) \) is a function of bounded terms, implying that \( \alpha_2(t) \) is also bounded. This ensures that \( x_3 = z_3 + \alpha_2 \) is bounded as well.
Following this line of argument, for \( i = 3, \dots, n-1 \), the stabilizing function \( \alpha_i(t) \) depends on bounded signals, as shown in equation \eqref{alphai}, ensuring the boundedness of \( \alpha_i(t) \) and the state \( x_{i+1}(t) \). 

Since \( u = \alpha_n \) from equation \eqref{alphai}, and \( \alpha_n(t) \) is bounded, the control input \( u(t) \) is also bounded. Therefore, all closed-loop signals are bounded.

(iv)
From (i) and (iii), the signals \( z_i(t) \), \( i = 1, \dots, n \), are bounded, and consequently, \( V_n(t) \) remains bounded. Given that \( \dot{V}_n(t) \) is non-increasing and bounded, it implies that \( \ddot{V}_n(t) \) is bounded as well. By applying Barbalat’s Lemma, we conclude that \( \dot{V}_n(t) \to 0 \) as \( t \to \infty \), which leads to \( z_i(t) \to 0 \), for \( i = 1, \dots, n \), ensuring the output tracking error \( z_1(t) \) converges to zero asymptotically.

\subsection{p-BLF for Higher-Order Nonlinear Systems with Full State Constraints}

Denote $z_1 = x_1 - x_d$ and $z_i = x_i - \alpha_{i-1}$ for $i = 2, \dots, n$. Consider the Lyapunov function candidate:
\begin{equation} \label{vful}
V = \sum_{i=1}^{n} V_i, \quad V_i = \frac{1}{2\beta} \ln \left( \frac{k_{i}^2}{k_{i}^2 - z_i^2} \right), \quad i = 1, \dots, n
\end{equation}

\noindent
where $k_1=k_{x1}-k_{xd}$, and $k_i$, for $i = 2, \dots, n$, are positive constants, and  $\beta > 0$ is a design parameter. $V$ is positive definite and continuously differentiable in the set $|z_i| < k_i$ for all $i = 1, 2, \dots, n$.

The stabilizing functions and control law are designed as:

\begin{equation} \label{fullalpha1}
\alpha_1 = \frac{1}{g_1} \left( -f_1 - \beta \kappa_1 z_1 \left( k_1^2 - z_1^2 \right) + \dot{y}_d \right)
\end{equation}

\begin{equation} \label{fullalphai}
 \begin{split}
\begin{aligned} 
\alpha_i = \frac{1}{g_i} \bigg( -f_i + \dot{\alpha}_{i-1} - \beta \kappa_i z_i \left( k_i^2 - z_i^2 \right) \\
- \frac{g_{i-1} z_{i-1}}{\beta} \frac{k_i^2 - z_i^2}{k_{i-1}^2 - z_{i-1}^2} \bigg), \quad i = 2, \dots, n
\end{aligned}
\end{split}
\end{equation}

\begin{equation} \label{fullu}
u = \alpha_n
\end{equation}
This yields the closed-loop system:
\begin{equation} \label{fullz1}
\dot{z}_1 = -\beta \kappa_1 z_1 \left( k_1^2 - z_1^2 \right) + g_1 z_2
\end{equation}

\begin{equation} \label{fullzi}
\begin{split}
\dot{z}_i &= -\beta \kappa_i z_i \left( k_i^2 - z_i^2 \right) 
- \frac{g_{i-1} z_{i-1}}{\beta} \frac{k_i^2 - z_i^2}{k_{i-1}^2 - z_{i-1}^2} \\
&+ g_i z_{i+1}, \quad i = 2, \dots, n-1
\end{split}
\end{equation}

\begin{equation}  \label{fullzn}
\dot{z}_n = -\beta \kappa_n z_n \left( k_n^2 - z_n^2 \right) - \frac{g_{n-1} z_{n-1}}{\beta} \frac{k_n^2 - z_n^2}{k_{n-1}^2 - z_{n-1}^2}
\end{equation}

The time derivative of $V$ can be rewritten as:
\begin{equation}  \label{fullvdot}
\dot{V} = -\sum_{j=1}^{n} \kappa_j z_j^2
\end{equation}

\textbf{Theorem 2:}  
Consider the nonlinear system described by \eqref{fullz1}–\eqref{fullzn} with the control law \eqref{fullu} under Assumption 1, and assume the initial conditions satisfy  \( |z_i(0)| < k_i \) for \( i = 1, 2, \dots, n \). Let the design parameters \( \beta > 0 \) and \( \kappa_i > 0 \) be chosen appropriately. Then, the closed-loop system ensures the following properties:

(i) The signals \( z_i(t) \), \( i = 1, \dots, n \), remain bounded within the compact set \( \Omega_z \), where \( D_{z_i} \) is defined as:
\[
D_{z_i} = k_i \sqrt{1 - e^{-2\beta V(0)} \prod_{j=1}^{n} \left( \frac{k_j^2 - z_j^2(0)}{k_j^2} \right)}
\]
for all \( i = 1, \dots, n \).

(ii) The full states \( x_i(t) \), \( i = 1, \dots, n \), remain within the constraint set \( |x_i(t)| < k_{xi} \) for all \( t \geq 0 \).

(iii) All closed-loop signals are bounded.

(iv) The tracking error \( z_i(t) = x_i(t) - x_d(t) \) converges to zero asymptotically as \( t \to \infty \), i.e., \( \lim_{t \to \infty} z_1(t) = 0 \), and the states \( z_i(t) \to 0 \), for \( i = 2, \dots, n \).

\textbf{Proof:}

i) Since \( \dot{V} \leq 0 \), we have \( V(t) \leq V(0) \) for all \( t \geq 0 \), implying:
\[
\frac{1}{2 \beta} \ln \left( \frac{k_i^2}{k_i^2 - z_i^2} \right) \leq \sum_{j=1}^{n} \frac{1}{2 \beta} \ln \left( \frac{k_j^2}{k_j^2 - z_j^2(0)} \right)
\]
Thus, \( |z_i(t)| \leq D_{z_i} \) holds for all \( t \geq 0 \), where \( i = 1, \dots, n \).

(ii) From the boundedness of \( z_1(t) \) in (i), we have:
\(
|x_1(t)| = |z_1(t) + x_d(t)| \leq D_{z_1} + |x_d(t)|
\).
Since \( |x_d(t)| \leq k_{xd} \), it follows that:
\(
|x_1(t)| \leq D_{z_1} + k_{xd} = k_{x1}
\).
For \( i = 2, \dots, n \), since \( x_i(t) = z_i(t) + \alpha_{i-1}(t) \), the boundedness of \( z_i(t) \) and \( \alpha_{i-1}(t) \) (from (iii) below) ensures that the states \( x_i(t) \) remain bounded within the constraint set \( |x_i(t)| < k_{xi} \).

(iii) From (i), the error signals \( z_1(t), \dots, z_n(t) \) are bounded. Considering Assumption 1, and the boundedness of \( z_1(t) \)  imply that \( x_1(t) \) is also bounded. From \eqref{fullalpha1}, the stabilizing function \( \alpha_1(t) \) is bounded, and thus, \( x_2 = z_2 + \alpha_1 \) is bounded. Similarly, from \eqref{fullalphai}, the stabilizing functions \( \alpha_i(t) \), for \( i = 2, \dots, n-1 \), are bounded due to their dependence on bounded terms, ensuring that the states \( x_i(t) \), \( i = 2, \dots, n \), are bounded. Since \( u = \alpha_n \) from \eqref{fullu}, and \( \alpha_n(t) \) is bounded, the control input \( u(t) \) is also bounded. Therefore, all closed-loop signals are bounded.

(iv) From (i) and (iii), the signals \( z_i(t) \), \( i = 1, \dots, n \), are bounded, which implies that \( V(t) \) is bounded. Given that \( \dot{V}(t) \leq 0 \) is non-increasing, it follows that \( \ddot{V}(t) \) is bounded. By applying Barbalat’s Lemma, we conclude that \( \dot{V}(t) \to 0 \) as \( t \to \infty \), leading to \( z_i(t) \to 0 \) for \( i = 1, \dots, n \), ensuring that the tracking error \( z_1(t) \to 0 \) asymptotically as \( t \to \infty \).

\section{Simulation}

We consider the second-order nonlinear system:
\[
\dot{x}_1 = \Theta_1 x_1^2 + x_2, \quad \dot{x}_2 = \Theta_2 x_1 x_2 + \Theta_3 x_1 + (1 + x_1^2) u,
\]

\noindent
where \(\Theta_1 = 0.1\), \(\Theta_2 = 0.1\), and \(\Theta_3 = -0.2\) \cite{tee2009barrier}. The control design employs a p-BLF for output constraints, with \(k_1 = 0.56\). The objective is to track the desired trajectory \(y_d = 0.2 + 0.3 \sin(t)\), ensuring that the state \(x_1\) remains within the safe constraint \(|x_1| < k_1\).
The initial conditions are set as \(x_1(0) = 0.25\) and \(x_2(0) = 1.5\). The control parameters \(\kappa_1 = 2\) and \(\kappa_2 = 2\) are used to adjust the response, while \(\beta = 10\) governs the control effort transition near the boundaries.

\begin{figure}[ht]
\centering
\includegraphics[width=0.40\textwidth]{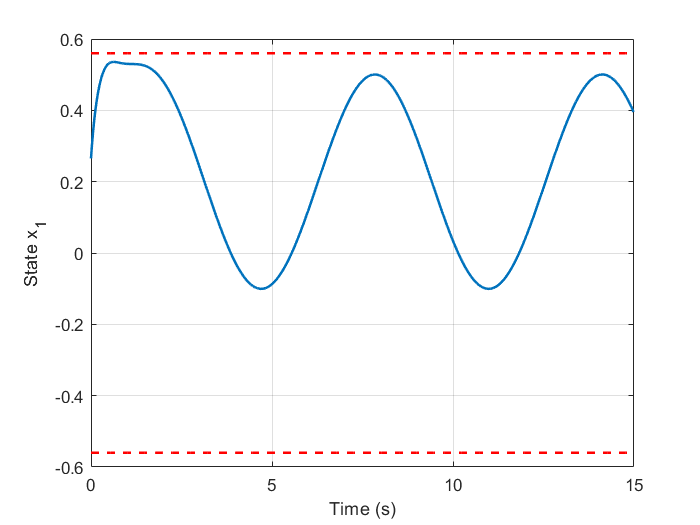}
\caption{State \(x_1\) over time with constraints \(|x_1| < 0.56\).}
\label{fig1}
\end{figure}

\begin{figure}[ht]
\centering
\includegraphics[width=0.40\textwidth]{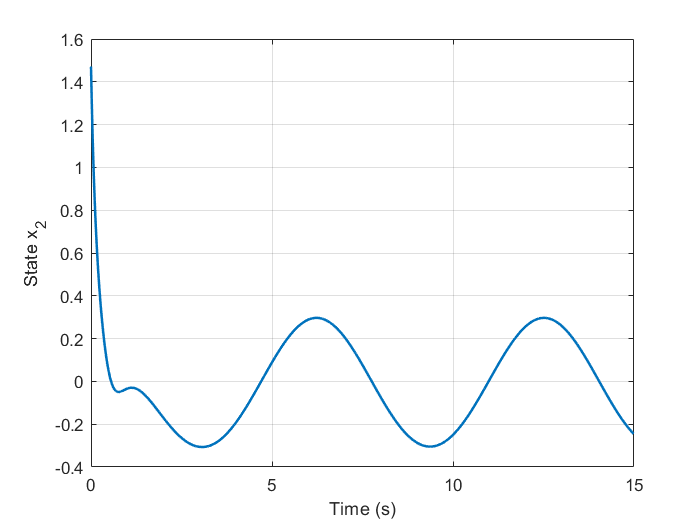}
\caption{State \(x_2\) over time.}
\label{fig2}
\end{figure}

\begin{figure}[ht]
\centering
\includegraphics[width=0.40\textwidth]{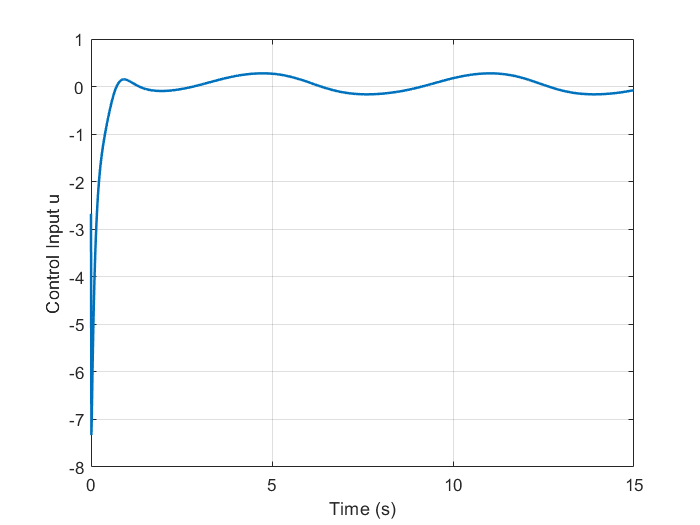}
\caption{Control input \(u\) over time.}
\label{fig3}
\end{figure}

\begin{figure}[ht]
\centering
\includegraphics[width=0.40\textwidth]{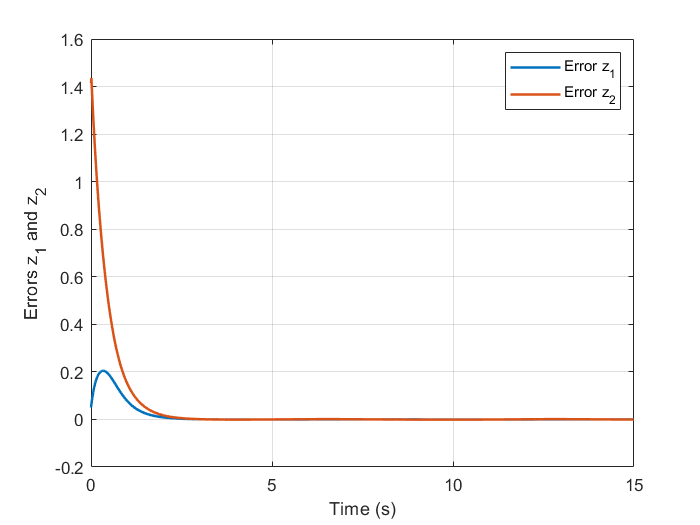}
\caption{Comparison of errors \(z_1\) and \(z_2\) over time.}
\label{fig4}
\end{figure}

\begin{figure}[ht]
\centering
\includegraphics[width=0.40\textwidth]{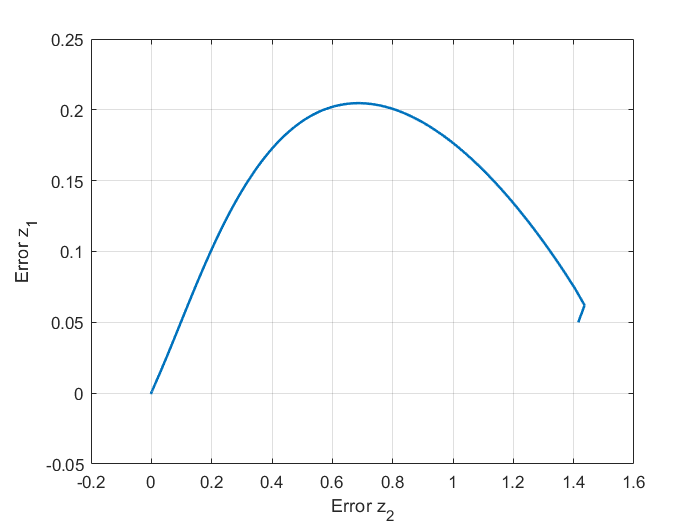}
\caption{Relationship between errors \(z_1\) and \(z_2\).}
\label{fig5}
\end{figure}

Figures \ref{fig1} and \ref{fig2} show the states \(x_1\) and \(x_2\) trajectories over time, with \(x_1\) constrained by \(|x_1| < 0.56\). Figure \ref{fig3} illustrates the control input \(u\) applied to the system. Figure \ref{fig4} compares the errors \(z_1\) and \(z_2\), while Figure \ref{fig5} presents the relationship between \(z_1\) and \(z_2\). The results demonstrate that the control law ensures constraint satisfaction and smooth system behavior.

\section{Conclusion}
In this paper, we introduced the p-BLF for output and full-state constrained nonlinear control systems. The method offers a novel approach by maintaining minimal control effort near the origin and smoothly increasing it as the system state moves toward the constraint boundaries, providing a continuous and stable control response. This gradual transition avoids abrupt changes in control effort, addressing limitations found in earlier methods, and improves system performance by reducing chattering. Two specific forms—logarithmic-based and rational-based—were proposed, each demonstrating their effectiveness through theoretical analysis and simulation results for output-constrained systems.
Future work will focus on optimizing the parameter $\beta$ to fine-tune the rate of control effort increase as the system state approaches the constraints. Additionally, we will explore the extension of p-BLF to multi-input multi-output systems and adaptive control schemes, as well as its application to systems with partial, asymmetric, time-varying, or unknown constraints to further enhance its practical applicability.

 \bibliographystyle{IEEEtran}

\end{document}